# Quasi-Molecular mechanism of cosmological recombination: a scheme of calculation

Tamaz Kereselidze[1], Irakli Noselidze[2] and Zaal Machavariani[1,3]

[1] Faculty of Exact and Natural Sciences, Tbilisi State University, Chavchavadze Avenue 3, 0179 Tbilisi, Georgia
[2] School of Science and Technology, University of Georgia, Kostava Str. 77a, 0171 Tbilisi, Georgia
[3] Doctoral School, Kutaisi International University, Youth Avenue, 5th Lane, 4600 Kutaisi, Georgia

**ABSTRACT**

For a quasi-molecular mechanism of cosmological recombination, a scheme of calculation based on a rigorous quantum-mechanical approach is elaborated. The probability of free-bound radiative transition into an excited state of a quasi-molecule temporarily formed by a colliding electron and two nearest neighboring protons is derived in a closed algebraic form.

## 1. INTRODUCTION

In our recent papers (Kereselidze, Noselidze & Ogilvie 2019, 2021, 2022 a, b) a non-standard quasi-molecular mechanism of recombination (QMR) was suggested and applied for the formation of atomic hydrogen in the early Universe. According to the QMR, an electron collides with two protons situated one far from another, emits a photon, and creates the hydrogen molecular ion, $H_2^+$ in a highly excited state. Depending on whether $H_2^+$ is created in an attractive or repulsive state, the temporarily formed quasi-molecule descends into a low-lying state or dissociates into the excited hydrogen atom and proton.

Following the standard mechanism of recombination two charged particles, an electron, and a proton participated in the formation of atomic hydrogen. An electron and a proton combined efficiently into the hydrogen atom only in a highly excited state, from which a rapid cascade occurred into a state with the principal quantum number $n = 2$ (Peebles 1968; Zel'dovich, Kurt & Syunyaev 1968). The linear size $d_n$ of the hydrogen atom being in a highly excited state is large ($d_n \simeq 2n^2$). This fact justifies the assumption made by Kereselidze et al (2019), that the nearest neighboring proton participated in recombination in the pre-recombination stage of evolution of the Universe when the temperature and density of matter were higher than subsequently.

The influence of another proton on recombination decreases when the density of protons decreases, i.e. when the Universe expands and subsequently cools. Hence, the QMR transforms into the standard mechanism – the recombination of an electron on an isolated proton when temperature decreases. The standard mechanism of recombination is thus a limiting case of the QMR at small redshift $z$.

The presence of another proton reduces the symmetry of a field experienced by an electron involved in recombination from spherical to axial and leads to a Stark splitting of the hydrogen energy levels. These two effects lead in turn to radiative transitions that are forbidden in the recombination of an electron with an isolated proton. The participation of the nearest neighbouring proton in the process thus opens quasi-molecular channels and hence has an impact on the recombination history.

The present study has two purposes: for the QMR to elaborate a scheme of calculation based on a rigorous quantum-mechanical approach, and to obtain the probability of free-bound radiative transitions in a closed algebraic form. The problem is treated in an adiabatic representation. Unless otherwise indicated, atomic units ($e = m_e = \hbar = 1$) are used throughout the paper.

## 2. WAVEFUNCTIONS OF AN ELECTRON

A precise quantum-mechanical calculation of cosmological recombination requires a knowledge of the correct wavefunctions of an electron involved in the process in both the initial continuous and final discrete states. From a wave-mechanical point of view, the problem is to obtain the correct wavefunctions that are solutions of the Schrödinger equation, which in the adiabatic approximation reads

$$\left(-\frac{1}{2}\Delta - \frac{1}{r_a} - \frac{1}{r_b}\right)\Psi^{(\pm)} = \varepsilon\Psi^{(\pm)}. \tag{1}$$

Here, $r_a$ and $r_b = |\vec{r}_a - \vec{R}|$ are the distances from an electron to protons $a$ and $b$, respectively, $R$ is the distance between protons, $\varepsilon$ is the electron energy, and $\Psi^{(\pm)}$ denotes the two-centre wavefunction that is either symmetric ($+$) or antisymmetric ($-$) with respect to a reflection in the plane normal to and bisecting axis $\vec{R}$. For a free electron $\varepsilon = k^2/2 > 0$ and $\Psi^{(\pm)} \equiv \Psi_{\vec{k}}^{(\pm)}$; for a bound electron $\varepsilon \equiv \varepsilon(R) < 0$ and $\Psi^{(\pm)} = \Psi_{H_2^+}^{(\pm)}$.

Solutions of equation (1) are obtained mostly in numerical form. An application of numerical wavefunctions to cosmological recombination requires formidable computational efforts and is time-consuming. The problem hence requires an alternative treatment. Our purpose is thus to find solutions to equation (1) in closed algebraic forms that are applicable to cosmological recombination. For this, we make use of the Coulomb-Born approximation (for the continuous spectrum) and the asymptotic method (for the discrete spectrum).

3.1 Continuous spectrum wavefunction

To begin, we find the continuous spectrum wavefunction. At large distances between protons, the solution of equation (1) is expressible as

$$\Psi_{\vec{k}}^{(\pm)} = 2^{-1/2}\left(\Psi_{\vec{k}}^a \pm \Psi_{\vec{k}}^b\right). \tag{2}$$

Here, $\Psi_{\vec{k}}^a$ ($\Psi_{\vec{k}}^b$) is the wavefunction of an electron moving in the Coulomb field of proton $a$ ($b$) and that is perturbed by another remote proton, $\vec{k}$ is the wave-vector of a colliding electron that is moving along axis $z$ of the coordinate system. For definiteness we find wavefunction, $\Psi_{\vec{k}}^a$.

In the Coulomb-Born approximation, wavefunction $\Psi_{\vec{k}}^a$ that satisfies the appropriate boundary condition becomes written as

$$\Psi_{\vec{k}}^a(\vec{r}_a) = \Psi_{\vec{k}}^{(0)}(\vec{r}_a) + \int G^{(+)}(\vec{r}_a, \vec{r}_a')\left|\vec{r}_a' - \vec{R}\right|^{-1}\Psi_{\vec{k}}^{(0)}(\vec{r}_a')d\vec{r}_a'. \tag{3}$$

In (3), $G^{(+)}(\vec{r}_a, \vec{r}_a')$ is the Coulomb Green's function (CGF) and $\Psi_{\vec{k}}^{(0)}(\vec{r}_a)$ is the solution of equation (1) without the Coulomb field of proton $b$. Symbol ($+$) denotes an outgoing wave when $r \to \infty$. In the following equations, the lower index is omitted for coordinates $\vec{r}_a$ and $\vec{r}_a'$.

The CGF can be constructed from its spectral representation, in which the summation runs over the complete set of discrete and continuum eigenstates. Blinder (1981) showed that a summation explicitly written in terms of discrete and continuous eigenstates in parabolic coordinates leads to the integral representation of the CGF. Making use of the scheme of calculation developed by Blinder, we evaluate the CGF in the form convenient for our purpose (Kereselidze, Noselidze & Ogilvie 2022 a):

$$G^{(+)}(\vec{r},\vec{r}\,') = -\frac{ik}{2\pi} \sum_{\bar{m}=-\infty}^{\infty} e^{i\bar{m}(\varphi-\varphi')} \int_0^{\infty} e^{i\frac{k}{2}(\mu+\nu+\mu'+\nu')\cosh s} \\ \times \chi(s) J_{\bar{m}}\left(b(\mu\mu')^{1/2}\right) J_{\bar{m}}\left(-b(\nu\nu')^{1/2}\right) ds. \quad (4)$$

In (4) $\mu = r(1+\cos\vartheta)$, $\nu = r(1-\cos\vartheta)$, $\varphi = \arctan(y/x)$ are parabolic coordinates, in which $r$ is the radial variable, $\vartheta$ is the polar angle and $\varphi$ is the azimuthal angle; $\chi(s) = \sinh s \left(\coth(s/2)\right)^{2i/k}$, $b = k\sinh s$ and $J_{\bar{m}}(x)$ is a Bessel function.

Expanding $\left|\vec{r}\,' - \vec{R}\right|^{-1}$ in terms of spherical harmonics, we write that

$$\frac{1}{\left|\vec{r}\,' - \vec{R}\right|} = \sum_{l,m} \frac{4\pi}{2l+1} \begin{Bmatrix} r'^l / R^{l+1} \\ R^l / r'^{l+1} \end{Bmatrix}_{r'>R}^{r'<R} Y_{lm}^*(\vartheta_{\vec{R}}, \varphi_{\vec{R}}) Y_{lm}(\vartheta', \varphi'). \quad (5)$$

For $R \gg 1$ and $r' \sim 1$, in (5) the first term is $O(R^{-1})$, the second is $O(R^{-2})$ and so on, whereas all terms are $O(R^{-1})$ when $r' \sim R$.

Wavefunction $\Psi_{\vec{k}}^{(0)}(\vec{r}\,')$ is oscillatory with decreasing amplitude as $r'$ increases. Because $R$ is extremely large, we can assume that the contribution of the space where $r' > R/2$ into the integral in (3) is small. This fact allows us to restrict the integration to $r'_{max} = R/2$ in (3) and write that

$$\Psi_{\vec{k}}(\vec{r}) = \Psi_{\vec{k}}^{(0)}(\vec{r}) + \frac{1}{R} \sum_{l,m} \frac{(l-|m|)!}{(l+|m|)!} P_l^{|m|}(\vartheta_{\vec{R}}) \\ \cdot \int G^{(+)}(\vec{r},\vec{r}\,')(r'/R)^l P_l^{|m|}(\vartheta') e^{im\varphi'} \Psi_{\vec{k}}^{(0)}(\vec{r}\,') d\vec{r}\,'. \quad (6)$$

In (6) $P_l^m(\vartheta_{\vec{R}})$ is a Legendre polynomial, $\vartheta_{\vec{R}}$ is the angle between $\vec{R}$ and axis $z$; angles $\varphi$ and $\varphi_{\vec{R}}$ are measured from plane $(\vec{z},\vec{R})$, therefore $\varphi_{\vec{R}} = 0$ in (6).

Inserting (4) in (6) and performing the integration over $\varphi'$, we obtain for the wavefunction of a colliding electron that

$$\Psi_{\vec{k}} = \Psi_{\vec{k}}^{(0)}(\mu,\nu) + \frac{1}{R}\Psi_{\vec{k}}^{(1)}(\mu,\nu,\varphi) + O(R^{-2}). \quad (7)$$

Here, the second term is a correction to the unperturbed wavefunction $\Psi_{\vec{k}}^{(0)}$, in which

$$\Psi_{\vec{k}}^{(1)} = -\frac{ik}{4} \sum_{l,m} \frac{(l-|m|)!}{(l+|m|)!} P_l^{|m|}(\cos\vartheta_{\vec{R}}) A_{l,m}(\mu,\nu) e^{im\varphi}, \quad (8)$$

and

$$A_{l,m} = \frac{1}{(2R)^l} \int_0^\infty \chi(s) \int_0^{\mu'_{max}} \int_0^{\nu'_{max}} e^{ik\cosh s(\mu+\nu+\mu'+\nu')/2} P_l^{|m|}\left((\mu'-\nu')/(\mu'+\nu')\right) \qquad (9)$$
$$\cdot J_m\left(b(\mu\mu')^{1/2}\right) J_m\left(-b(\nu\nu')^{1/2}\right) \Psi_{\vec{k}}^{(0)}(\mu',\nu')(\mu'+\nu')^{l+1} d\mu' d\nu' ds.$$

In (9), the upper limits of integration are chosen from the condition $\mu'_{max} + \nu'_{max} = R$ that corresponds to $r'_{max} = R/2$. The explicit expression of $\Psi_{\vec{k}}^{(0)}$ is presented in appendix A. Wavefunction $\Psi_{\vec{k}}^b$ is defined with the above equations in which $\mu_a \to \nu_b$ and $\nu_a \to \mu_b$.

2.2 Discrete spectrum wavefunction

Protons might be located arbitrarily with respect to the direction of propagation of a colliding electron. This condition means that angle $\vartheta_{\vec{R}}$ that defines orientation of $\vec{R}$ with respect to axis $z$ can take values from zero to $\pi/2$. Let $(\tilde{x}, \tilde{y}, \tilde{z})$ be a rotating coordinate system with axis $\tilde{z}$ along $\vec{R}$, while $(x, y, z)$ is a fixed coordinate system with axis $z$ along $\vec{k}$.

Equation (1) is separable in prolate spheroidal coordinates $\tilde{\xi} = (\tilde{r}_a + \tilde{r}_b)/R$ ($1 \le \tilde{\xi} < \infty$), $\tilde{\eta} = (\tilde{r}_a - \tilde{r}_b)/R$ ($-1 \le \tilde{\eta} \le 1$) and $\tilde{\varphi} = arctg(\tilde{y}/\tilde{x})$ ($0 \le \tilde{\varphi} < 2\pi$). Representing wavefunction $\Psi^{(\pm)}$ as this product function

$$\Psi = \frac{U_1(\tilde{\xi})}{(\tilde{\xi}^2-1)^{1/2}} \frac{U_2(\tilde{\eta})}{(1-\tilde{\eta}^2)^{1/2}} \Phi_{\tilde{m}}(\tilde{\varphi}), \qquad (10)$$

in which $\Phi_{\tilde{m}} = e^{\pm i\tilde{m}\tilde{\varphi}}/\sqrt{2\pi}$, and inserting (10) into (1), we obtain two equations for unknown functions

$$U_1'' + \left[-\frac{\gamma^2 R^2}{4} + \frac{\lambda + 2R\tilde{\xi}}{\tilde{\xi}^2 - 1} + \frac{1-\tilde{m}^2}{(\tilde{\xi}^2-1)^2}\right] U_1 = 0,$$
$$U_2'' + \left[-\frac{\gamma^2 R^2}{4} - \frac{\lambda}{1-\tilde{\eta}^2} + \frac{1-\tilde{m}^2}{(1-\tilde{\eta}^2)^2}\right] U_2 = 0. \qquad (11)$$

Here, $\lambda$ is the separation constant depending on $R$ and $\gamma = \sqrt{-2\varepsilon(R)} = \sqrt{1/n^2 + 2/R + O(R^{-2})}$ Bates & Reid (1968); $\tilde{m}$ denotes the absolute value of the projected orbital angular momentum of an electron along molecular axis $\tilde{z}$. Functions $U_1(\tilde{\xi})$ and $U_2(\tilde{\eta})$ satisfy the boundary conditions: $U_1(1) = U_1(\infty) = 0$ and $U_2(\pm 1) = 0$.

When $R$ tends to infinity an electron might be attached to either proton. To consider both these possibilities, we introduce new variables as follows:

$$\rho_1 = \gamma R(\tilde{\xi} - 1), \qquad \rho_2 = \gamma R(1 \pm \tilde{\eta}),$$
$$0 \le \rho_1 < \infty, \qquad 0 \le \rho_2 \le 2\gamma R. \qquad (12)$$

In (12), sign ($+$) corresponds to the case in which, as $R \to \infty$, an electron is attached to proton $a$ that is located on the molecular axis with coordinates $\tilde{\xi} = 1$, $\tilde{\eta} = -1$ (left center); sign ($-$) corresponds to the case in which, as $R \to \infty$, an electron is attached to proton $b$ that is located on axis $\tilde{z}$ with coordinates $\tilde{\xi} = 1$, $\tilde{\eta} = 1$ (right center).

In new variables equations (11) become expressible as

$$U_{1,2}'' + \left[ -\frac{1}{4} + \left( \frac{1+\tau_{1,2} \pm \lambda/R}{2\gamma} \mp \frac{1-\tilde{m}^2}{4\gamma R} \right) \frac{1}{\rho_{1,2}} + \frac{1-\tilde{m}^2}{4\rho_{1,2}^2} \right.$$

$$\pm \left( \frac{1+\tau_{1,2} \mp \lambda/R}{4\gamma^2 R} \pm \frac{1-\tilde{m}^2}{8\gamma^2 R^2} \right) \frac{1}{1 \pm \rho_{1,2}/2\gamma R} \quad (13)$$

$$\left. + \frac{1-\tilde{m}^2}{16\gamma^2 R^2} \frac{1}{(1 \pm \rho_{1,2}/2\gamma R)^2} \right] U_{1,2} = 0,$$

in which $\tau_{1,2} = \pm 1$. When $R$ is much larger than the linear size of an electron shell on nucleus ($R \gg d_n$), ratios $\rho_1/R$ and $\rho_2/R$ are small quantities in the main region of distribution of an electron. Making use of these small parameters the asymptotically exact solutions of equations (13) have been found (Kereselidze, Noselidze & Chibisov 2003). These solutions are

$$U_{1,2}(\rho_{1,2}) = e^{-\frac{\alpha_{1,2}\rho_{1,2}}{2}} \rho_{1,2}^{\frac{\tilde{m}+1}{2}} F\left(-n_{1,2}, \tilde{m}+1, \alpha_{1,2}\rho_{1,2}\right). \quad (14)$$

Here, $F(-n_i, \tilde{m}+1, \alpha_i\rho_i)$ is a confluent hypergeometric function, $n_1$, $n_2$, $\tilde{m}$ are parabolic quantum numbers that are related to the principle quantum number by equation $n = n_1 + n_2 + \tilde{m} + 1$, and $\alpha_{1,2} = \left[ 1 - n^2 R^{-1} \left( 1 + \tau_{1,2} - \lambda R^{-1} \right) \right]^{1/2}$, in which $\lambda R^{-1} = n^{-1}(n_1 - n_2) - 1$.

For $R \gg r_n$, $\rho_1 \simeq \gamma(1 + \tilde{\mu}/2R)\tilde{\nu}$ and $\rho_2 \simeq \gamma(1 - \tilde{\nu}/2R)\tilde{\mu}$ (Bates & Reid 1968) where $\tilde{\mu} = \tilde{r}(1 + \cos\tilde{\vartheta})$, $\tilde{\nu} = \tilde{r}(1 - \cos\tilde{\vartheta})$, $\tilde{\varphi} = \arctan(\tilde{y}/\tilde{x})$ are parabolic coordinates with origin in the left centre. In these coordinates wavefunction (10) reads

$$\Psi_{n_1 n_2 \tilde{m}} = C_{n_1 n_2 \tilde{m}} e^{-\frac{\gamma}{2}(\alpha_1 \tilde{\nu} + \alpha_2 \tilde{\mu})} (\tilde{\nu}\tilde{\mu})^{\frac{\tilde{m}}{2}} \left(1 - \frac{\tilde{\nu}-\tilde{\mu}}{2R}\right)^{\frac{\tilde{m}+1}{2}}$$

$$\cdot F\left(-n_1, \tilde{m}+1, \gamma\alpha_1\left(1+\frac{\tilde{\mu}}{2R}\right)\tilde{\nu}\right) F\left(-n_2, \tilde{m}+1, \gamma\alpha_2\left(1-\frac{\tilde{\nu}}{2R}\right)\tilde{\mu}\right) \Phi_{\tilde{m}}(\tilde{\varphi}), \quad (15)$$

where $C_{n_1 n_2 \tilde{m}}(R)$ is a normalising factor. The region of applicability of wavefunction (15) is restricted by the conditions $\tilde{\mu} + \tilde{\nu} \leq R$.

Expanding $\alpha_1$, $\alpha_2$ and $\gamma$ over small parameter $1/R$, we can write that

$$\Psi_{n_1 n_2 \tilde{m}} = \Psi_{n_1 n_2 \tilde{m}}^{(0)}(\tilde{\mu}, \tilde{\nu}, \tilde{\varphi}) + \frac{1}{R}\Psi_{n_1 n_2 \tilde{m}}^{(1)}(\tilde{\mu}, \tilde{\nu}, \tilde{\varphi}) + O(R^{-2}). \quad (16)$$

In (16), $\Psi_{n_1 n_2 \tilde{m}}^{(0)}$ is the wavefunction of the hydrogen atom in parabolic coordinates, and $R^{-1}\Psi_{n_1 n_2 \tilde{m}}^{(1)}$ is its correction. The explicit expressions of $\Psi_{n_1 n_2 \tilde{m}}^{(0)}$ and $\Psi_{n_1 n_2 \tilde{m}}^{(1)}$ are presented in appendix A.

Parabolic coordinates $\tilde{\mu}$, $\tilde{\nu}$ and $\tilde{\varphi}$ are related to parabolic coordinates of the fixed system according to equations

$$\tilde{\mu} = \frac{\mu+\nu}{2} + \frac{\mu-\nu}{2}\cos\vartheta_{\vec{R}} + \sqrt{\mu\nu}\cos\varphi\sin\vartheta_{\vec{R}},$$

$$\tilde{\nu} = \frac{\mu+\nu}{2} - \frac{\mu-\nu}{2}\cos\vartheta_{\vec{R}} - \sqrt{\mu\nu}\cos\varphi\sin\vartheta_{\vec{R}}, \quad (17)$$

$$\tilde{\varphi} = \arctan \frac{\sqrt{\mu\nu}\sin\varphi}{\sqrt{\mu\nu}\cos\varphi\cos\vartheta_{\vec{R}} - \frac{\mu-\nu}{2}\sin\vartheta_{\vec{R}}}.$$

We thus obtain that the wavefunction centred on proton $a$, as $R \to \infty$, is defined with equation (16), in which parabolic coordinates are given according to (17). The wavefunction centred on proton $b$, as $R \to \infty$, is defined with the above equations in which $\mu_a \to \nu_b$ and $\nu_a \to \mu_b$.

## 3. SCHEME OF CALCULATION

In the fixed coordinate system, the operator of electric-dipole strength is $\vec{d} = -(\vec{i}x + \vec{j}y + \vec{k}z)$. For convenience, we calculate the matrix elements of operators $d^{(\pm)} = -(x \pm iy)$ and $d^{(z)} = -z$. In parabolic coordinates these operators read $d^{(\pm)} = -\sqrt{\mu\nu}e^{\pm i\varphi}$ and $d^{(z)} = -(\mu-\nu)/2$.

The probability of a radiative transition depends on distance $R$ between protons, and is defined as (Heitler 1954)

$$W_{i,f}(R) = \frac{4\omega_{if}^3(R)}{3c^3}\left|\vec{d}_{i,f}(R)\right|^2. \tag{18}$$

In (18), $\omega_{if}$ is the frequency of an emitted photon, $c$ is the speed of light, and $\vec{d}_{i,f}$ is the transition matrix element.

The free-bound transition probability calculated over wavefunctions (7) and (16) can be represented as

$$W_{i,f}(R) = W_{i,f}^{(0)} + \frac{1}{R}W_{i,f}^{(1)} + O(R^{-2}), \tag{19}$$

In (19), $W_{i,f}^{(0)}$ determines the transition probability on an isolated proton,

$$W_{i,f}^{(1)} = \frac{4\omega_0^2}{c^3}\left|\left\langle \Psi_{n_1 n_2 \tilde{m}}^{(0)}\left|\vec{d}\right|\Psi_{\vec{k}}^{(0)}\right\rangle\right|^2 + \frac{8\omega_0^3}{3c^3}\mathrm{Re}\Bigg[\left\langle \Psi_{n_1 n_2 \tilde{m}}^{(0)}\left|\vec{d}\right|\Psi_{\vec{k}}^{(0)}\right\rangle \\ \cdot \left(\left\langle \Psi_{n_1 n_2 \tilde{m}}^{(1)}\left|\vec{d}\right|\Psi_{\vec{k}}^{(0)}\right\rangle + \left\langle \Psi_{n_1 n_2 \tilde{m}}^{(0)}\left|\vec{d}\right|\Psi_{\vec{k}}^{(1)}\right\rangle\right)^*\Bigg], \tag{20}$$

accounts for a correction caused by the participation of another proton in recombination, and $\omega_{if}^{(0)} = 2^{-1}(k^2 + n^{-2})$ is the main term of $\omega_{if}(R)$; $W_{i,f}^{(0)}$ is determined by equations (7), (16) and (18) in which $R = \infty$.

The QMR is distinguished from the standard mechanism of recombination by the presence of strongly attractive terms of $H_2^+$ at large $R$ with $m=0$ and $n_1 < n_2$ (blue curves in Fig. 1 in Kereselidze et al 2021). To calculate transition probabilities into these states, we treat the case when protons are oriented parallel with respect to the direction of propagation of a colliding electron ($\vartheta_{\vec{R}} = 0$). In this case, $d_{n_1 n_2 0, k}^{(\pm)} \equiv 0$, and hence, the problem is reduced to the calculation of matrix elements $d_{n_1 n_2 0, k}^{(z)}$ in (20).

To begin, we calculate matrix element $\left\langle \Psi_{n_1 n_2 0}^{(1)}\left|d^{(z)}\right|\Psi_{\vec{k}}^{(0)}\right\rangle$ in (20). Performing an appropriate calculation, we obtain that

$$\left\langle \Psi^{(1)}_{n_1 n_2 0} \middle| d^{(z)} \middle| \Psi^{(0)}_{\vec{k}} \right\rangle = \frac{\sqrt{2\pi} C^{(0)}_{n_1 n_2 0} C_k}{16} \left[ n_1(2n_2+1)\left(f^{(3,0)}_{n_1-1,n_2} - f^{(1,2)}_{n_1-1,n_2}\right) \right.$$

$$-n_2(2n_1+1)\left(f^{(2,1)}_{n_1,n_2-1} - f^{(0,3)}_{n_1,n_2-1}\right) - \frac{n_1}{n}\left(f^{(3,1)}_{n_1-1,n_2} - f^{(1,3)}_{n_1-1,n_2}\right)$$

$$+\frac{n_2}{n}\left(f^{(3,1)}_{n_1,n_2-1} - f^{(1,3)}_{n_1,n_2-1}\right) + n_2\left(f^{(3,0)}_{n_1,n_2} - f^{(1,2)}_{n_1,n_2}\right) - n_1\left(f^{(2,1)}_{n_1,n_2} - f^{(0,3)}_{n_1,n_2}\right)$$

$$\left. + \frac{2 C^{(1)}_{n_1 n_2 0}}{C^{(0)}_{n_1 n_2 0}}\left(f^{(2,0)}_{n_1,n_2} - f^{(0,2)}_{n_1,n_2}\right) \right], \quad (21)$$

in which

$$f^{(p_1,p_2)}_{n_1,n_2-1} = \int_0^R e^{-a_1 \nu} F\left(-n_1, 1, \frac{\nu}{n}\right) F\left(\frac{i}{k}, 1, ik\nu\right) f^{(p_2)}_{n_2-1}(\nu) \nu^{p_1} d\nu,$$

$$f^{(p_1,p_2)}_{n_1-1,n_2} = \int_0^R e^{-a_1 \nu} F\left(-(n_1-1), 2, \frac{\nu}{n}\right) F\left(\frac{i}{k}, 1, ik\nu\right) f^{(p_2)}_{n_2}(\nu) \nu^{p_1} d\nu, \quad (22)$$

$$f^{(p_2)}_{n_2} = \int_0^{R-\nu} e^{-a_2 \mu} F(-n_2, 1, \mu/n) \mu^{p_2} d\mu,$$

$$f^{(p_2)}_{n_2-1} = \int_0^{R-\nu} e^{-a_2 \mu} F(-(n_2-1), 2, \mu/n) \mu^{p_2} d\mu, \quad (23)$$

and $a_{1,2} = (1/n \pm ik)/2$.

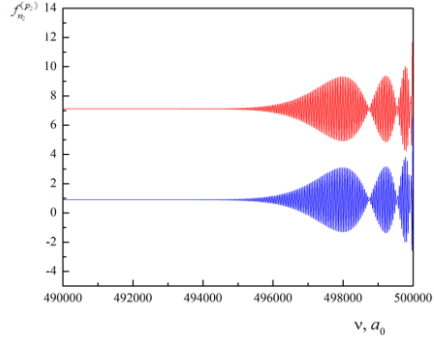

Fig.1 The behavior of $f^{(p_2)}_{n_2}$ at large $\nu$. The real part – blue curve, the imaginary part – red curve; $z = 3000$, $R = 10^5 a_0$, $p_2 = 0$ and $n = 3$. Similar behavior is observable for $p_2 = 1, 2, 3$ and $n > 3$.

In Fig. 1 $f^{(p_2)}_{n_2}$ is depicted as a function of $\nu$. The data are obtained by performing a numerical integration in (23). The figure clearly shows that $f^{(p_2)}_{n_2}$ is a constant at all $\nu$, except the region near $R$ where it is strongly oscillatory. Assuming that $f^{(p_2)}_{n_2}$ is a constant at all $\nu$, we can write that

$$f^{(p_1,p_2)}_{n_1,n_2} = f^{(p_2)}_{n_2} \int_0^{R-\Delta} e^{-a_1 \nu} F\left(-n_1, 1, \frac{\nu}{n}\right) F(i/k, 1, ik\nu) \nu^{p_1} d\nu, \quad (24)$$

where

$$f_{n_2}^{(p_2)} = \sum_{j=0}^{n_2} \frac{(-1)^j d_j}{(j!)^2 n^j} \frac{(p_2 + j)!}{a_2^{p_2+j+1}}, \qquad (25)$$

and $\Delta$ is chosen from the condition that in (22) the integral from $R-\Delta$ to $R$ is less than a one-ten thousandth of the integral from zero to $R-\Delta$. We note that (25) is obtained by analytical integration of (23), in which $F(-n_2,1,v/n)$ is represented in a polynomial form with coefficients $d_0 = 1$, $d_1 = n_2$, $d_2 = n_2(n_2-1)$, $d_3 = n_2(n_2-1)(n_2-2)$, $\cdots d_{n_2} = n_2!$. For $f_{n_2-1}^{(p_2)}$ $n_2 \to n_2 - 1$ and $(j!)^2 \to (j+1)! j!$ in (25).

In (24) $F(-n_1,1,v/n)$ increases as $(v/n)^{n_1}$ when $v$ increases. On the other hand, as the real part of $a_1$ is positive, $e^{-a_1 v}$ rapidly decreases when $v$ increases. Thus, in (24) the integrand exponentially decreases when $v$ increases. This fact allows us to extend the integration up to infinity in (24). In this case, this integral becomes analytically solvable. Assuming that $R = \infty$ and $p_1 = 0$ in (24), we obtain that (Landau & Lifshitz 1977)

$$f_{n_1,i/k}^{(0)} = \int_0^\infty e^{-a_1 v} F(-n_1,1,v/n) F(i/k,1,ikv) dv$$
$$= \frac{2n}{(1+ink)^{n_1+1-i/k}} \frac{(ink-1)^{n_1}}{(1-ink)^{i/k}} F\left(-n_1, \frac{i}{k}, 1, -\frac{4ink}{(1-ink)^2}\right), \qquad (26)$$

in which $F(-n_1, i/k, 1, x)$ is a hypergeometric function. Integrals with $p_1 > 0$ can be reduced to (26) by making use of the recurrence relation

$$f_{n_1,i/k}^{(p_1)} = \frac{4n}{1+(nk)^2}\left[\left(\frac{1-ink}{2} + n_1 - n - i(p_1-1)nk\right)f_{n_1,i/k}^{(p_1-1)}\right.$$
$$\left. + n(p_1-1)\left(p_1 - 1 - \frac{2i}{k}\right)f_{n_1,i/k}^{(p_1-2)} + \frac{2in(p_1-1)}{k}f_{n_1,i/k+1}^{(p_1-2)}\right]. \qquad (27)$$

We thus obtain that $f_{n_1 n_2 0}^{(p_1,p_2)}$ is the product of two quantities $f_{n_2}^{(p_2)}$ and $f_{n_1,i/k}^{(p_1)}$ that are defined with equations (25)-(27). The data obtained by equations (25)-(27) agree well with the results of numerical calculations.

In (20), matrix element $\langle \Psi_{n_1 n_2 0}^{(0)} | d^{(z)} | \Psi_{\vec{k}}^{(1)} \rangle$ may be evaluated by changing the order of integration to perform the integration over $\mu$, $v$, $\varphi$ first, and then over $\mu'$, $v'$, $s$

$$\langle \Psi_{n_1 n_2 0}^{(0)} | d^{(z)} | \Psi_{\vec{k}}^{(1)} \rangle = -\frac{i\pi^{1/2} k C_k}{16 n^2} \sum_{l=0}^{\infty} \frac{(-1)^l}{(2R)^l} \int_0^\infty \chi(s) \int_0^R \int_0^{R-v'} e^{\frac{ik}{2}(1+\cosh s)\mu'} e^{-\frac{ik}{2}(1-\cosh s)v'}$$
$$\cdot \left(g_{n_1 n_2 0}^{(2,0)} - g_{n_1 n_2 0}^{(0,2)}\right) F(i/k,1,ikv') P_l\left((\mu'-v')/(\mu'+v')\right)(\mu'+v')^{l+1} d\mu' dv' ds. \qquad (28)$$

Here,

$$g_{n_1 n_2 0}^{(p_1,p_2)}(\mu',v') = \int_0^R e^{-av} J_0(b\sqrt{v'v}) F\left(-n_1,1,\frac{v}{n}\right) g_{n_2}^{(p_2)}(\mu',v) v^{p_1} dv, \qquad (29)$$

in which

$$g_{n_2}^{(p_2)} = \int_0^{R-v} e^{-a\mu} J_0(b\sqrt{\mu'\mu}) F\left(-n_2,1,\frac{\mu}{n}\right) \mu^{p_2} d\mu, \qquad (30)$$

and $a = (1/n - ik\cosh s)/2 > 0$.

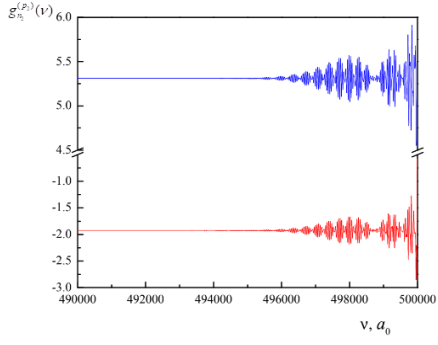

Fig.2 The behavior of $g_{n_2}^{(p_2)}$ at large $v$. The real part – blue curve, the imaginary part – red curve; $z = 3000$, $R = 10^5 a_0$, $p_2 = 0$, $n = 3$ and $b(\mu')^{1/2} = 1$. Similar behavior is observable for $p_2 = 2$, $n > 3$ and $b(\mu')^{1/2} > 1$.

Numerical calculation shows that $g_{n_2}^{(p_2)}$ is constant at all $v$, except the region near $R$ where it is strongly oscillatory (see Fig. 2). Assuming that $g_{n_2}^{(p_2)}$ is constant at all $v$, we obtain that

$$g_{n_1 n_2 0}^{(p_1, p_2)} = g_{n_2}^{(p_2)}(\mu') \int_0^{R-\Delta} e^{-av} J_0\left(b\sqrt{v'v}\right) F\left(-n_1, 1, \frac{v}{n}\right) v^{p_1} dv, \tag{31}$$

where

$$g_{n_2}^{(p_2)} = \sum_{i=0}^{\infty} \sum_{j=0}^{n_2} \frac{(-1)^{i+j} b^{2i} d_j}{2^{2i} (i!)^2 n^j (j!)^2} \frac{(p_2 + i + j)!}{a^{p_2+i+j+1}} \mu'^i. \tag{32}$$

As in the previous case, in (31) $\Delta$ is chosen from the condition that the integral from $R-\Delta$ to $R$ is less than a one-ten thousandth of the integral from zero to $R-\Delta$. We note that (32) is obtained by the expansion of $J_0\left(b\sqrt{v'v}\right)$ in (30) and subsequent analytical integration.

Increasing the upper limit of integration to infinity and representing $F(-n_1, 1, v/n)$ in a polynomial form, the integral in (31) becomes analytically solvable (Gradshtein & Ryzhik 1980)

$$g_{n_1}^{(p_1)}(v') = \int_0^{\infty} e^{-av} J_0\left(b\sqrt{v'v}\right) F\left(-n_1, 1, \frac{v}{n}\right) v^{p_1} dv$$
$$= \sum_{j=0}^{n_1} \frac{(-1)^j d_j}{(j!)^2 n^j} \frac{(j+p_1)!}{a^{j+p_1+1}} F\left(j+p_1+1, 1, -\frac{b^2 v'}{4a}\right). \tag{33}$$

Because in (33) hypergeometric function $F(j+p_1+1, 1, -b^2 v'/4a)$ is a product of an exponential function and a polynomial of order $j+p_1$, (28) reduces to the solution of integrals of this type

$$\int_0^R e^{-\left[\frac{b^2}{4a} + \frac{ik}{2}(1-\cosh s)\right]v'} F\left(\frac{i}{k}, 1, ikv'\right) v'^{k_1} \int_0^{R-y} e^{\frac{ik}{2}(1+\cosh s)\mu'} \mu'^{k_2} d\mu' dv', \tag{34}$$

where $k_1$ and $k_2$ are integers. These integrals are analytically solvable. The integral over s, we calculate numerically in (28).

The derived equations allow us to calculate the probability of free-bound radiative transition into an arbitrary excited state of $H_2^+$ with $m = 0$.

# APPENDIX A

In parabolic coordinates

$$\Psi_{\vec{k}}^{(0)} = C_k e^{i\frac{k}{2}(\mu-\nu)} F(i/k, 1, ik\nu), \quad (A1)$$
$$C_k = (2\pi)^{-3/2} e^{\pi/2k} \Gamma(1-i/k),$$

where $F(i/k, 1, ik\nu)$ is a confluent hypergeometric function, $\Gamma(1-i/k)$ is a gamma function and $c_k$ is a normalising factor.

The normalized wavefunction of the hydrogen atom in parabolic coordinates is

$$\Psi_{n_1 n_2 \tilde{m}}^{(0)} = C_{n_1 n_2 \tilde{m}}^{(0)} e^{-\frac{\tilde{\mu}+\tilde{\nu}}{2n}} (\tilde{\mu}\tilde{\nu})^{\frac{\tilde{m}}{2}} F\left(-n_1, \tilde{m}+1, \frac{\tilde{\nu}}{n}\right) F\left(-n_2, \tilde{m}+1, \frac{\tilde{\mu}}{n}\right) \Phi_{\tilde{m}}(\tilde{\varphi}), \quad (A2)$$

$$C_{n_1 n_2 \tilde{m}}^{(0)} = \frac{2^{1/2}}{n^2 n^{\tilde{m}} (\tilde{m}!)^2} \sqrt{\frac{(n_1+\tilde{m})!}{n_1!} \frac{(n_2+\tilde{m})!}{n_2!}}.$$

The explicit expression of $\Psi_{n_1 n_2 \tilde{m}}^{(1)}$ reads

$$\Psi_{n_1 n_2 \tilde{m}}^{(1)} = \left\{ \frac{1}{2(m+1)} \left[ n_1 \left( 2n_2 + \tilde{m} + 1 - \frac{\tilde{\mu}}{n} \right) \frac{F(-(n_1-1), \tilde{m}+2, \tilde{\nu}/n)}{F(-n_1, \tilde{m}+1, \tilde{\nu}/n)} \tilde{\nu} \right. \right.$$
$$\left. - n_2 \left( 2n_1 + \tilde{m} + 1 - \frac{\tilde{\nu}}{n} \right) \frac{F(-(n_2-1), \tilde{m}+2, \tilde{\mu}/n)}{F(-n_2, \tilde{m}+1, \tilde{\mu}/n)} \tilde{\mu} \right] \quad (A3)$$
$$\left. - \frac{n_1 \tilde{\mu}}{2} + \frac{n_2 \tilde{\nu}}{2} + \frac{C_{n_1 n_2 \tilde{m}}^{(1)}}{C_{n_1 n_2 \tilde{m}}^{(0)}} \right\} \Psi_{n_1 n_2 \tilde{m}}^{(0)}(\tilde{\mu}, \tilde{\nu}, \tilde{\varphi}).$$

For $\tilde{m} = 0$ the normalising factor is

$$C_{n_1 n_2 0} = \frac{2\gamma^{3/2} \alpha_1 \alpha_2}{\sqrt{(2n_1+1)\alpha_2 + (2n_2+1)\alpha_1}} \simeq \frac{\sqrt{2}}{n^2}\left[1 + \frac{n(n_1-n_2)}{2R} + O(R^{-2})\right]. \quad (A4)$$